\documentclass{optica-article}

\journal{opticajournal} 

\articletype{Research Article}
\usepackage{fancyhdr}
\usepackage{lineno}
\usepackage{bbm}
\usepackage{mathrsfs}
\usepackage{amsmath}

\begin{document}

\title{Generation of optomicrowave and optomagnonic entanglements in cascaded optomagnomechanical systems}

\author{Chong Zhang,\authormark{1,2} Xiaomin Liu,\authormark{1,2} Liwen Gao,\authormark{1,2} Rongguo Yang,\authormark{1,2,3,4} Jing Zhang,\authormark{1,2,3,*} and Tiancai Zhang\authormark{1,2,3}}

\address{\authormark{1}College of Physics and Electronic Engineering, Shanxi University, Taiyuan 030006, China\\
\authormark{2}State Key Laboratory of Quantum Optics Technologies and Devices, Shanxi University, Taiyuan 030006, China\\
\authormark{3}Collaborative Innovation Center of Extreme Optics, Shanxi University, Taiyuan 030006, China\\
\authormark{4}yrg@sxu.edu.cn}
\email{\authormark{*}zjj@sxu.edu.cn}

\begin{abstract*} The optomagnomechanical system, which involves flexible nonlinearities, is one of the promising physical platforms for studying the preparation and manipulation of quantum entanglements, as well as the construction of hybrid quantum networks. A scheme for entanglement enhancement and quadripartite entanglement generation is proposed, based on a cascaded optomagnomechanical system. On the one hand, optomicrowave and optomagnonic entanglements within the two subsystems are investigated, and their parameter dependence, such as detuning, decay, coupling strength, and transmission efficiency, is discussed. On the other hand, the parameter conditions for achieving optimal optomicrowave and optomagnonic quadripartite entanglements are also obtained. The results show that significant enhancement of optomicrowave and optomagnonic entanglements in the second cavity can be obtained in a certain range of parameters. Under optimized parameter conditions, optomicrowave and optomagnonic quadripartite entanglements can be generated throughout the entire cascaded system. This research provides a theoretical basis for the manipulation of quantum entanglement, the transmission of the magnon's state, and the construction of hybrid quantum networks involving different physical systems.
\end{abstract*}

\section{Introduction}
The preparation of quantum entanglement, especially multipartite quantum entanglement, is a prerequisite for realizing various quantum tasks, including quantum information processing\cite{419326112,Clerk2020HybridQS}, quantum network construction\cite{Pirandola2016PhysicsUT,062620}, quantum computation\cite{Ladd2010QuantumC,Du23}, and quantum simulation\cite{153}. Usually, the implementation of different quantum tasks requires different physical systems. For example, superconducting qubits running in the microwave frequency band have advantages in quantum information processing\cite{Arute2019QuantumSU}, while optical signals propagating through optical fibers facilitate the long-distance transmission of quantum information\cite{260502,Xavier2019QuantumIP}. Therefore, the connection among different quantum systems is necessary for building hybrid quantum networks. The phonon and magnon modes in optomagnomechanical system can couple with electromagnetic fields of almost any frequency, making it possible to obtain entanglement between different physical systems. In other words, the optomagnomechanical system is an ideal platform for generating hybrid quantum entanglement (especially optomicrowave and optomagnonic entanglements) as well as for constructing hybrid quantum networks\cite{Fan2022MicrowaveOpticsEV,Luo23,Di23,023501}. An optomagnomechanical system composed of a microwave cavity, an optical cavity, and a microbridge-structured YIG crystal attached with a small high-reflectivity mirror pad, was proposed and studied, while the steady optomicrowave entanglement can be achieved if the optical cavity is driven by a red-detuned laser field and the magnon mode is driven by a blue-detuned microwave field \cite{Fan2022MicrowaveOpticsEV}. Considering this system under a PT-symmetric-like condition with microwave cavity gain, enhanced optomicrowave, optomagnonic, and microwave-phonon entanglements were obtained\cite{Luo23}. Driving two identical hybrid optomagnomechanical systems with a two-mode squeezed optical field, the entanglement between optical modes was first transferred to the phonon modes via optomechanical coupling, then to the magnon modes via magnomechanical coupling (magnetostriction interaction), and finally to the two microwave modes through magnetic dipole interactions\cite{Di23}. In the hybrid optomagnomechanical system, in which a two-level atomic ensemble is added, the optomechanical entanglement generated by the optomechanical coupling was transferred to the atom and magnon modes through magnomechanical and Tavis-Cummings (TC) couplings, that is, the atomic-magnon entanglement was obtained\cite{023501}. Unfortunately, the generated hybrid quantum entanglements are often fragile due to quantum decoherence and environmental perturbations. Therefore, enhancing the entanglement is becoming a critical task. Cascading is one of the effective methods of entanglement enhancement and has been widely adopted in various physical systems\cite{Xin17,040305}. Using a cascaded non-degenerate optical parametric amplifier system, the entanglement between signal and idler photons generated in the parametric down conversion process can be significantly enhanced\cite{040305}. Similarly, cascading two four-wave mixing systems resulted in obvious enhancement of entanglement between the probe and conjugate lights\cite{Xin17}. In this work, we propose a scheme for enhancing both optomicrowave and optomagnonic entanglements, as well as generating quadripartite entanglement, based on a cascaded optomagnomechanical system. We find the optimal parameters of entanglement enhancement and quadripartite entanglement generation, and give the basis of the chosen parameters. This research is valuable for the construction of hybrid quantum networks and offers new insights into the manipulation and utilization of the magnon's quantum properties.

\begin{figure}[htbp]
\centering\includegraphics[width=13cm]{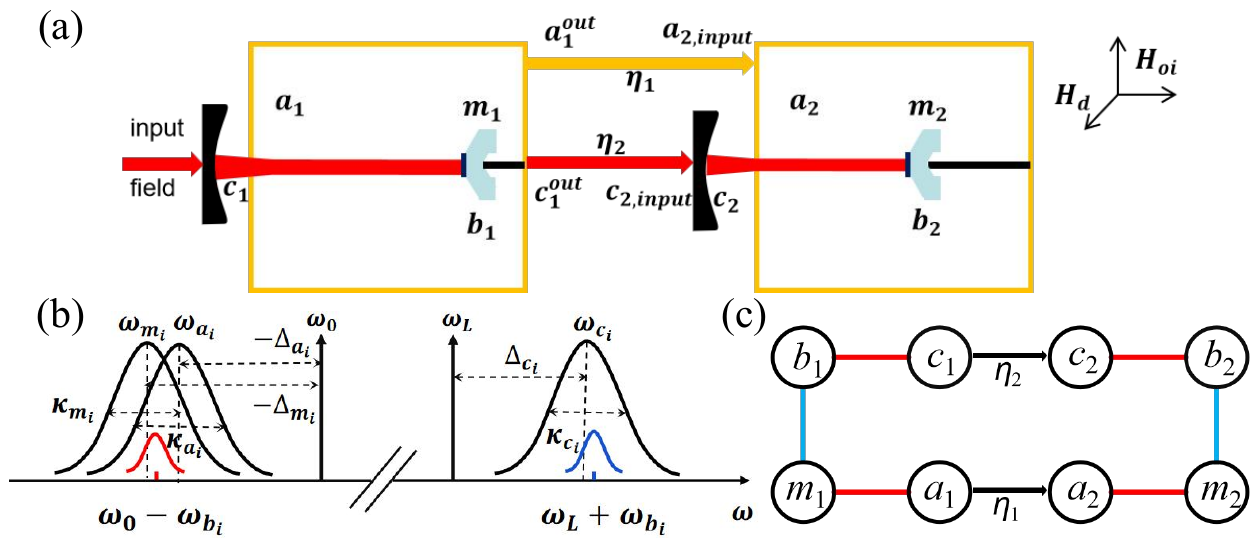}
\caption{(a) Schematic diagram of the cascaded optomagnomechanical system for generating optomagnonic and optomicrowave entanglements. (b) frequency relation. (c) interaction between various modes.}
\end{figure}
The case in which two optomagnomechanical systems are cascaded is under consideration, as shown in Figure 1(a), where each optomagnomechanical system includes a microwave mode $a_i$, a magnon mode $m_i$, a phonon mode $b_i$, and an optical mode $c_i$, $i=1,2$. The optical cavity consists of a cavity mirror and a microbridge-structured YIG crystal attached with a small high-reflectivity mirror pad. In the YIG crystal, not only the magnon mode $m_i$ can be induced upon a uniform bias magnetic field and a microwave driving field, but also the phonon modes $b_i$ can be generated due to the magnetostrictive effect. The YIG crystal is strategically placed in the maximum magnetic region of the microwave mode, ensuring strong magnetic-dipole interactions between the magnon and microwave modes. The magnon mode couples with the phonon mode via magnetostriction interaction, while the phonon mode couples with the optical modes through radiation pressure interaction. The frequency relations of various modes are depicted in Fig.1(b). In optomagnomechanical system 1, the magnon mode $m_1$ is driven by a blue-detuned microwave field, which means the magnon frequency is resonant with the Stokes sideband of the driving microwave field, $\omega_{m_i} \simeq \omega_0-\omega_{b_i}$, where $\omega_{m_i}$, $\omega_0$ and $\omega_{b_i}$ are the frequency of the magnon mode, the driven microwave field, and the phonon mode, respectively. In this case, the interaction between the magnon mode and the phonon mode is dominated by the two-mode squeezing mechanism, which facilitates the generation of entanglement between the magnon and phonon modes (magnon-phonon entanglement)\cite{031201,0370,070101}. The microwave frequency is also near the Stokes sideband of the driving microwave field. Simultaneously, the optical mode is driven by a red-detuned laser, which is resonant with the anti-Stokes sideband of the driving field, i.e., $\omega_{c_1} \simeq \omega_L+\omega_{b_1}$, where $\omega_{c_1}$ and $\omega_{L}$ are the frequency of the first cavity mode and the driven optical field, respectively. Thus, the interaction between the optical mode and the phonon mode is dominated by a beam-splitter (state transfer) interaction mechanism\cite{030405,30005,29581}, which means that the generated magnon-phonon entanglement can be transferred to the optical modes. Since the magnetic dipole interaction between the magnon and microwave modes is also a beam-splitter (state-swap) interaction mechanism, the generated magnon-phonon entanglement can also be transferred to the microwave modes. As a result, optomagnonic and optomicrowave entanglements can be generated. The output fields from the optical and microwave cavities in optomagnomechanical system 1 are subsequently fed into the optical and microwave cavities of optomagnomechanical system 2, forming a cascaded system, where the interaction between various modes is shown in Fig.1(c).
In the rotating frame of the driven fields, the Hamiltonian can be expressed as
\begin{equation}
\begin{aligned}
H/\hbar =& \sum_{i=1}^2( \Delta_{a_i} a_i^{\dagger} a_i + \Delta_{m_i} m_i^{\dagger} m_i + \Delta_{c_i} c_i^{\dagger} c_i + \frac{\omega_{b_i}}{2} (q_i^2 + p_i^2)\\
&+ g_{am} (a_i^{\dagger}m_i + a_i m_i^{\dagger}) + g_{m_ib_i}m_i^{\dagger}m_iq_i - g_{c_ib_i}c_i^{\dagger}c_iq_i\\ &+ i\Omega_i(m_i^\dagger - m_i) + iE(c_1^\dagger - c_1),
\end{aligned}
\end{equation}
The four terms of the first row of Eq.(1) are the free Hamiltonians of the microwave, magnon, optical, and phonon modes, respectively, where $o_i=a_i,m_i,c_i,b_i$($o_i^\dagger$) are the annihilation (creation) operators of microwave photon, magnon, optical photon, and phonon, satisfying the commutation relations $[o_i,o_i^\dagger]=1$. $q_i$ and $p_i$ are the dimensionless position and momentum of the mechanical vibration mode, $b=(q+ip)/\sqrt{2}$. $\Delta_{a_i}=\omega_{a_i}-\omega_0$, $\Delta_{m_i}=\omega_{m_i}-\omega_0$, and $\Delta_{c_i}=\omega_{c_i}-\omega_L$ are the detunings of the microwave mode, the magnon mode, and the optical mode, respect to their driving fields. $\omega_{m_i}=\gamma H_{0i}$ is the frequency of the magnon, $\gamma$ is the gyromagnetic ratio, and $H_{0i}$ is the amplitude of the bias magnetic field. The three terms in the second row represent the magnetic dipole, magnetostrictive and radiation pressure interactions, with $g_{am}$, $g_{m_ib_i}$, and $g_{c_ib_i}$ being the corresponding coupling strengths. The two terms in the third row are the microwave and laser driving terms, with the Rabi frequency $\Omega=\frac{\sqrt{5}}{4}\gamma\sqrt{N}H_d$ representing the coupling strength between magnon mode and the microwave driving field\cite{203601}, where $N$ is the total number of spins and $H_d$ is the amplitude of the driving magnetic field. $E=\sqrt{2\kappa_{c_1}p_L\hbar\omega_L}$ is the coupling strength between the optical mode and the driving laser, where $P_L$ and $\omega_L$ are the power and frequency of the laser, and $\kappa_{c_1}$ is the decay rate of optical mode 1. 

According to the cascade system shown in Fig.1(a), Considering the output microwave and optical modes $a_1^{out}(t)$ and $c_1^{out}(t)$ of system 1 as the input microwave and optical modes $a_{2,input}(t)$ and $c_{2,input}(t)$ of system 2, and taking into account the microwave and optical transmission efficiencies between two systems, $\eta_1$ and $\eta_2$, the input modes of system 2 can be expressed as\cite{013192}
\begin{equation}
\begin{aligned}
a_{2,input} &= \sqrt{\eta_{1}} a_{1}^{out}(t) + \sqrt{1-\eta_{1}} a_2^{in}(t) \\
c_{2,input} &= \sqrt{\eta_{2}} c_{1}^{out}(t) + \sqrt{1-\eta_{2}} c_2^{in}(t)
\end{aligned}
\end{equation}
where $a_1^{out}(t)$ and $c_1^{out}(t)$ are the output microwave and optical modes of system 1, $a_2^{in}$ and $c_2^{in}$ are the input noise operators of the microwave and optical modes of system 2, respectively. Taking into account the input-output relations $a_1^{out}(t) = \sqrt{\kappa_{a_1}} a_1(t) + a_1^{in}(t), c_1^{out}(t) = \sqrt{\kappa_{c_1}} c_1(t) + c_1^{in}(t)$,
$a_{2,input}(t)$ and $c_{2,input}(t)$ can be further expressed as
\begin{equation}
\begin{aligned}
a_{2,input}(t) &= \sqrt{2\eta_{1} \kappa_{a_1}} a_1(t) + \sqrt{\eta_1} a_1^{in}(t) + \sqrt{1-\eta_1} a_2^{in} \\
c_{2,input}(t) &= \sqrt{2\eta_{2} \kappa_{c_1}} c_1(t) + \sqrt{\eta_2} c_1^{in}(t) + \sqrt{1-\eta_2} c_2^{in}
\end{aligned}
\end{equation}
Considering dissipation and input noise of each mode, the quantum Langevin equations of the system can be expressed as
\begin{equation}
\begin{aligned}
\dot{a}_1 &= -i\Delta_{a_1}a_1 -\kappa_{a_1} a_1 -ig_{am} m_1 -\sqrt{2\kappa_{a_1}} a_1^{in}(t) \\
\dot{m}_1 &= -i\Delta_{m_1}m_1 -\kappa_{m_1} m_1 -ig_{am} a_1 -ig_{m_1b_1}m_1q_1 +\Omega -\sqrt{2\kappa_{m_1}} m_1^{in}(t) \\
\dot{c}_1 &= -i\Delta_{c_1}c_1 - \kappa_{c_1} c_1 +ig_{c_1b_1} c_1q_1 + E - \sqrt{2\kappa_{c_1}} c_1^{in}(t) \\
\dot{q}_1 &= \omega_{b_1} p_1 \\
\dot{p}_1 &= -\omega_{b_1} q_1 - \gamma_{b_1} p_1 -g_{m_1b_1} m_1^\dagger m_1 +g_{c_1b_1}c_1^\dagger c_1 + \xi_1(t) \\
\dot{a}_2 &= -i\Delta_{a_2}a_2 - \kappa_{a_2} a_2 -ig_{am} m_2 -\sqrt{2\eta_1\kappa_{a_1}\kappa_{a_2}}a_1 - \sqrt{2\eta_1\kappa_{a_2}} a_1^{in} - \sqrt{2(1-\eta_1)\kappa_{a_2}} a_2^{in}(t) \\
\dot{m}_2 &= -i\Delta_{m_2}m_2 - \kappa_{m_2} m_2 -ig_{am} a_2 -ig_{m_2b_2}m_2q_2 +\Omega - \sqrt{2\kappa_{m_2}} m_2^{in}(t) \\
\dot{c}_2 &= -i\Delta_{c_2}c_2 - \kappa_{c_2} c_2 +ig_{c_2b_2} c_2q_2 -\sqrt{2\eta_2\kappa_{c_1}\kappa_{c_2}}c_1 - \sqrt{2\eta_2\kappa_{c_2}} c_1^{in} - \sqrt{2(1-\eta_2)\kappa_{c_2}} c_2^{in}(t) \\
\dot{q}_2 &= \omega_{b_2} p_2 \\
\dot{p}_2 &= -\omega_{b_2} q_2 - \gamma_{b_2} p_2 -g_{m_2b_2} m_2^\dagger m_2 +g_{c_2b_2}c_2^\dagger c_2 + \xi_2(t)
\end{aligned} 
\end{equation}
where $\kappa_{a_i}$, $\kappa_{m_i}$, $\kappa_{c_i}$, and $\gamma_{b_i}$ are the decay rates for the four modes, and $o_i^{in}$, ($o_i=a_i,m_i,c_i$) is the corresponding input noise operators. The noise correlation terms can be expressed as $\langle o_i^{in}(t) {o_i^{in}}^{\dagger}(t^\prime) \rangle = (\overline{n}_{o_i+1})\delta(t-t^{\prime})$ and $\langle {o_i^{in}}^{\dagger}(t) o_i^{in}(t^\prime) \rangle = \overline{n}_{o_i} \delta(t-t^\prime)$. 
$\xi_i(t)$ is the noise operator for the phonon mode, and its correlation function is given by $\langle \xi_i(t) \xi_i(t') + \xi_i(t') \xi_i(t) \rangle/2 \simeq \gamma_{b_i} (2\overline{n}_{b_i}+1) \delta(t-t^\prime)$, where $\overline{n}_{b_i} = [\exp(\hbar \omega_{b_i} / k_B T)- 1]^{-1}$ with $T$ and $k_B$ representing the environmental temperature and the Boltzmann constant. As shown in the frequency relationship in Fig.1(b), $|\Delta_{o_{i}}|\sim\omega_{b_{i}}$, $o_{i}=a_{i},m_{i},c_{i}$. Under the resolved sideband condition $\omega_{b_{i}}\gg\kappa_{o_{i}}$, the simplified steady-state solutions are given by $\langle a_1 \rangle \simeq -ig_{am}\Omega/(g_{am}^2-\widetilde{\Delta}_{m_1} \Delta_{a_1})$, $\langle c_1 \rangle=-iE/\widetilde{\Delta}_{c_1}$, $\langle c_2 \rangle=\sqrt{\eta_2} \langle c_1 \rangle/10$, $\langle m_1 \rangle=i\Omega\Delta_{a_1}/(g_{am}^2-\widetilde{\Delta}_{m_1} \Delta_{a_1})$, $\langle m_2 \rangle=i\Omega\Delta_{a_2}/(g_{am}^2-\widetilde{\Delta}_{m_2} \Delta_{a_2})$. Here, the effective detunings of the magnon and optical modes are given by $\widetilde{\Delta}_{m_{i}}=\Delta_{m_{i}}+g_{m_{i}b_{i}}\left\langle q_{i}\right\rangle$ and $\widetilde{\Delta}_{c_{i}}=\Delta_{c_{i}}-g_{c_{i}b_{i}}\langle q_{i}\rangle$ (including the frequency shifts induced by magnetostrictive interaction and radiation pressure effect). The effective magnon-phonon coupling strength is defined by $G_{m_ib_i}=-i\sqrt{2}g_{m_ib_i} \langle m_i \rangle$, while the effective optomechanical coupling strength is given by $G_{c_ib_i}=i\sqrt{2}g_{c_ib_i}\langle c_i \rangle$.

When both hybrid optomagnomechanical systems are driven strongly, each mode can be expressed as the sum of its mean value and fluctuation. And the fluctuation of each mode can be further represented by its quadrature amplitude operator $\delta X_{o_i}(t)$ and quadrature phase operator $\delta Y_{o_i}(t)$, satisfying $\delta X_{o_i}(t)=(\delta o_i+\delta o_i^\dagger)/\sqrt{2}$ and $\delta Y_{o_i}(t)=i(\delta o_i^\dagger-\delta o_i)/\sqrt{2}$, $\delta o_i=\delta a_i,\delta m_i, \delta c_i$. Finally, the linearized Langevin equations describing the time evolution of the quadrature operators can be expressed in matrix form as
\begin{equation}
\begin{aligned}
\dot{u}(t) &= \mathcal{A} x(t) + n(t),
\end{aligned} 
\end{equation}
where $u(t)=[\delta X_{a_1}, \delta Y_{a_1}, \delta X_{m_1}, \delta Y_{m_1}, \delta q_1, \delta p_1, \delta X_{c_1}, \delta Y_{c_1}, \delta X_{a_2}, \delta Y_{a_2}, \delta X_{m_2}, \delta Y_{m_2}, \delta q_2, \delta p_2, \\ \delta X_{c_2}, \delta Y_{c_2}]^T$ is a set of the quadrature fluctuations, and the $16 \times 16$ drift matrix $\mathcal{A}$ can be expressed by
\begin{equation}
\mathcal{A}=\begin{pmatrix}
 \mathcal{A}_{a_1} & \mathcal{A}_{am} & 0 & 0 & 0 & 0 & 0 & 0\\
 \mathcal{A}_{am} & \mathcal{A}_{m_1} & \mathcal{A}_{m_1b_1} & 0 & 0 & 0 & 0 & 0\\
 0 & \mathcal{A}_{m_1b_1}^* & \mathcal{A}_{b_1} & \mathcal{A}_{c_1b_1}^* & 0 & 0 & 0 & 0\\
 0 & 0 & \mathcal{A}_{c_1b_1} & \mathcal{A}_{c_1} & 0 & 0 & 0 & 0\\
 \mathcal{A}_{a_{12}} & 0 & 0 & 0 & \mathcal{A}_{a_2} & \mathcal{A}_{am} & 0 & 0\\
 0 & 0 & 0 & 0 &\mathcal{A}_{am} & \mathcal{A}_{m_2b_2}^* & \mathcal{A}_{b_2} & \mathcal{A}_{c_2b_2}^*\\
 0 & 0 & 0 &\mathcal{A}_{c_{12}} & 0 & 0 & \mathcal{A}_{c_2b_2} & \mathcal{A}_{c_2}\\
\end{pmatrix}
\end{equation}
where each element is a $2\times2$ block matrix and the non-zero block matrices are $\begin{aligned}
 \mathcal{A}_{a_i}=\begin{pmatrix}
 -\kappa_{a_i} & \Delta_{a_i}\\
 -\Delta_{a_i} & -\kappa_{a_i}\\
\end{pmatrix} 
\end{aligned}$, $\begin{aligned}
 \mathcal{A}_{m_i}=\begin{pmatrix}
 -\kappa_{m_i} & \widetilde{\Delta}_{m_i}\\
 -\widetilde{\Delta}_{m_i} & -\kappa_{m_i}\\
\end{pmatrix} 
\end{aligned}$, $\begin{aligned}
 \mathcal{A}_{b_i}=\begin{pmatrix}
 0 & \omega_{b_i}\\
 -\omega_{b_i} & -\gamma_{b_i}\\
\end{pmatrix} 
\end{aligned}$, $\begin{aligned}
 \mathcal{A}_{c_i}=\begin{pmatrix}
 -\kappa_{c_i} & \widetilde{\Delta}_{c_i}\\
 -\widetilde{\Delta}_{c_i} & -\kappa_{c_i}\\
\end{pmatrix} 
\end{aligned}$, $\begin{aligned}
 \mathcal{A}_{am}=\begin{pmatrix}
 0 & g_{am}\\
 -g_{am} & 0\\
\end{pmatrix} 
\end{aligned}$, $\begin{aligned}
 \mathcal{A}_{m_ib_i}=\begin{pmatrix}
 G_{m_ib_i} & 0\\
 0 & 0\\
\end{pmatrix} 
\end{aligned}$, $\begin{aligned}
 \mathcal{A}_{m_ib_i}^*=\begin{pmatrix}
 0 & 0\\
 0 & -G_{m_ib_i}\\
\end{pmatrix} 
\end{aligned}$, $\begin{aligned}
 \mathcal{A}_{c_ib_i}=\begin{pmatrix}
 G_{c_ib_i} & 0\\
 0 & 0\\
\end{pmatrix} 
\end{aligned}$, $\begin{aligned}
 \mathcal{A}_{c_ib_i}^*=\begin{pmatrix}
 0 & 0\\
 0 & -G_{c_ib_i}\\
\end{pmatrix} 
\end{aligned}$,\\ $\begin{aligned}
 \mathcal{A}_{a_{12}}=\begin{pmatrix}
 -2\sqrt{\eta_1\kappa_{a_1}\kappa_{a_2}} & 0\\
 0 & -2\sqrt{\eta_1\kappa_{a_1}\kappa_{a_2}}\\
\end{pmatrix} 
\end{aligned}$, $\begin{aligned}
 \mathcal{A}_{c_{12}}=\begin{pmatrix}
 -2\sqrt{\eta_2\kappa_{c_1}\kappa_{c_2}} & 0\\
 0 & -2\sqrt{\eta_2\kappa_{c_1}\kappa_{c_2}}\\
\end{pmatrix} 
\end{aligned}$.

The stability condition of the system requires that the real parts of all the eigenvalues of the drift matrix $\mathcal{A}$ are negative. The correlations between various modes of the system can be characterized by a 16$\times$16 covariance matrix whose element is defined by $V_{jk}=\langle u_j(t)u_k(t^\prime)+u_k(t^\prime)u_j(t) \rangle /2, (j,k=1,2,...,16)$. The steady-state covariance matrix can be obtained by solving the Lyapunov equation $\mathcal{A}V+V\mathcal{A}^T=-\mathcal{D}$, where the diffusion matrix $\mathcal{D}$ is defined by $D_{jk}\delta(t-t^{\prime})=\left\langle n_{j}(t)n_{k}(t^{\prime})+n_{k}(t^{\prime})n_{j}(t)\right\rangle/2$, and its matrix representation is
\begin{equation}
\begin{aligned}
\mathcal{D} = \begin{pmatrix}
\mathcal{D}_{a_1} & 0 & 0 & 0 & \mathcal{D}_{a_{12}} & 0 & 0 & 0\\
0 & \mathcal{D}_{m_1} & 0 & 0 & 0 & 0 & 0 & 0\\
0 & 0 & \mathcal{D}_{b_1} & 0 & 0 & 0 & 0 & 0\\
0 & 0 & 0 & \mathcal{D}_{c_1} & 0 & 0 & 0 & \mathcal{D}_{c_{12}}\\
\mathcal{D}_{a_{12}} & 0 & 0 & 0 & \mathcal{D}_{a_2} & 0 & 0 & 0\\
0 & 0 & 0 & 0 & 0 & \mathcal{D}_{m_2} & 0 & 0\\
0 & 0 & 0 & 0 & 0 & 0 & \mathcal{D}_{b_2} & 0\\
0 & 0 & 0 & \mathcal{D}_{c_{12}} & 0 & 0 & 0 & \mathcal{D}_{c_2}\\
\end{pmatrix}
\end{aligned}
\end{equation}
where each element is a block matrix $2\times2$ and the non-zero block matrices are $\mathcal{D}_{a_1}=\kappa_{a_1}(2\overline{n}_{a_1}+1)I$, $\mathcal{D}_{a_2}=\kappa_{a_2}[\eta_1(2\overline{n}_{a_1}+1)+(1-\eta_1)(2\overline{n}_{a_2}+1)]I$, $\mathcal{D}_{m_i}=\kappa_{m_i}(2\overline{n}_{m_i}+1)I$, $\mathcal{D}_{b_i}=diag(0, \gamma_{b_i}(2\overline{n}_{b_i}+1))$,
$\mathcal{D}_{c_1}=\kappa_{c_1}(2\overline{n}_{c_1}+1)I$,
$\mathcal{D}_{c_2}=\kappa_{c_2}[\eta_2(2\overline{n}_{c_1}+1)+(1-\eta_2)(2\overline{n}_{c_2}+1)]I$, $\mathcal{D}_{a_{12}}=\sqrt{\eta_1\kappa_{a_1}\kappa_{a_2}}I$, $\mathcal{D}_{c_{12}}=\sqrt{\eta_2\kappa_{c_1}\kappa_{c_2}}I$.

We adopt logarithmic negativity to quantify entanglement\cite{032314,090503}, defined as $E_N = \max[0, -\ln(2\eta^-)]$,
where $\eta^-=\min eig |\otimes_{j=1}^2(-\sigma_yPV_4P)|^2$ is the minimum symplectic eigenvalue of the partially transposed covariance matrix $PV_4P$, $\sigma_y$ is the Pauli matrix $y$, and $V_4$ is the covariance matrix containing the two modes of interest, and $P=diag(1,-1,1,1)$ denotes the partial transpose operation.

\begin{figure}[htbp]
\centering\includegraphics[width=13cm]{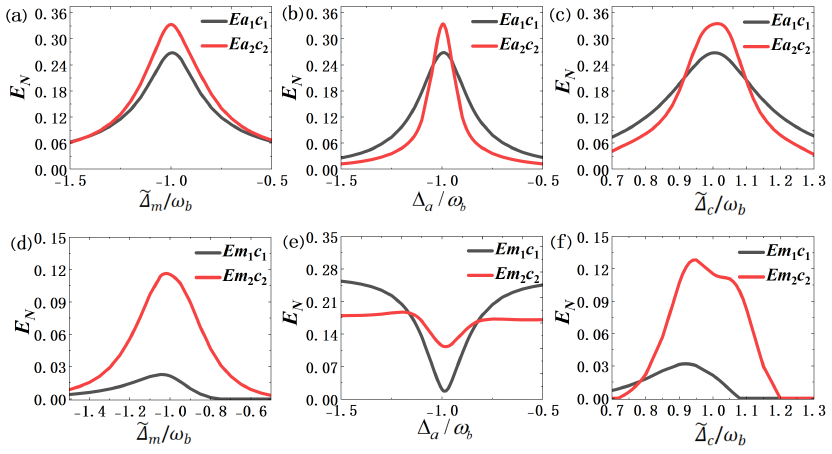}
\caption{The optomicrowave entanglements ($E_{a_1c_1}$ and $E_{a_2c_2}$) and the optomagnonic entanglements ($E_{m_1c_1}$ and $E_{m_2c_2}$) versus the effective detuning of magnon mode (a,d), the detuning of microwave mode (b,e), and the effective detuning of optical mode (c,f), respectively. The related parameters are: $\kappa_{a_i}=\kappa_a=2\pi\times1.5MHz$, $\kappa_{m_i}=\kappa_m=2\pi\times1.5MHz$, $\gamma_{b_i}=\gamma_b=2\pi\times100Hz$, $\kappa_{c_i}=\kappa_c=2\pi\times2MHz$, $\omega_{a_i}=\omega_a=2\pi\times10GHz$, $\omega_{m_i}=\omega_m=2\pi\times10GHz$, $\omega_{b_i}=\omega_b=2\pi\times40MHz$, $\lambda_{c}=1550nm$, $G_{c_1b_1}=2\pi\times8MHz$, $G_{m_ib_i}=2\pi\times2MHz$, $T=10mK$, $\eta_i=0.75$. $g_{am}=2\pi\times4MHz$ for subgraphs (a), (b) and (c); $g_{am}=2\pi\times6MHz$ for subgraphs (d), (e) and (f). The unchanged detunings are: $\Delta_{a_i}=\Delta_a=-\omega_b$, $\widetilde{\Delta}_{m_i}=\widetilde{\Delta}_m=-\omega_b$, $\widetilde{\Delta}_{c_i}=\widetilde{\Delta}_c=\omega_b$.}
\end{figure}
First, we consider the variation of the optomicrowave entanglement ($E_{a_1c_1}$ and $E_{a_2c_2}$) and optomagnonic entanglement ($E_{m_1c_1}$ and $E_{m_2c_2}$) of each system as a function of the respective detunings, as shown in Fig.2. Overall, compared to the optomicrowave and optomagnonic entanglements in system 1, the corresponding entanglements in system 2 can be significantly enhanced within certain parameter regions. The enhancement of entanglement is most pronounced near the strongest Stokes scattering point, and particularly the optomagnonic entanglement can be increased several times. This is because the cascaded process transmits the output fields, which carry the correlation information from the optical and microwave modes of system 1, into the optical and microwave modes of system 2, thereby enhancing the corresponding correlations in system 2. The enhancement of optomagnonic entanglement is not only from the correlations of system 1, but also from the entanglement generation process itself. As shown in Fig.2(a) and Fig.2(d), optomicrowave and optomagnonic entanglements can be enhanced over a broad range of magnon detuning. From Fig.2(b) and Fig.2(e), it can be observed that the enhancement of optomicrowave and optomagnonic entanglements are present only within a narrow range of microwave detuning, with the enhancement of optomagnonic entanglement occurring over a wider detuning range than optomicrowave entanglement. From Fig.2(c) and Fig.2(f), it can be seen that both the optomicrowave and optomagnonic entanglements are enhanced across a relatively wide range of optical detuning. In summary, the enhancement effect of entanglement is most sensitive to microwave detuning $\Delta_{a}$, followed by optical detuning $\widetilde{\Delta}_{c}$, and is least sensitive to magnon detuning $\widetilde{\Delta}_{m}$.

\begin{figure}[htbp]
\centering\includegraphics[width=13cm]{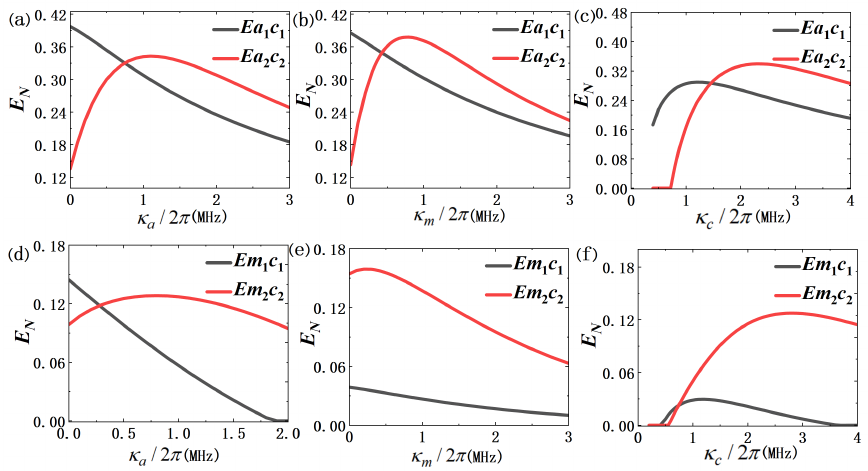}
\caption{The optomicrowave entanglement ($E_{a_1c_1}$ and $E_{a_2c_2}$) and the optomagnonic entanglement ($E_{m_1c_1}$ and $E_{m_2c_2}$) versus the decay of microwave mode (a,d), magnon mode (b,e), and optical mode (c,f), respectively. All parameters being consistent with those in Fig.2, except the variables.}
\end{figure}
How optomicrowave and optomagnonic entanglements vary with different decay rates for each system is shown in Fig.3. It can be seen that over a wide range of decay rates, both the optomicrowave entanglement $E_{a_2c_2}$ and the optomagnonic entanglement $E_{m_2c_2}$ in system 2 are significantly enhanced compared to the corresponding entanglements in system 1. Additionally, $E_{a_2c_2}$ and $E_{m_2c_2}$ exhibit a nonmonotonic trend (first increase, then decrease) with increasing losses. As shown in Fig.3(a), Fig.3(c), Fig.3(d), and Fig.3(f), when the decay rates $\kappa_{a}$ and $\kappa_{c}$ are very weak, no cascaded enhancement occurs. However, when the decay rates $\kappa_{a}$ and $\kappa_{c}$ are relatively strong, cascaded enhancement becomes noticeable, and the entanglement decreases with continuously increasing decay rates. This is because too weak decay rates $\kappa_{a}$ and $\kappa_{c}$ hinder the transmission of microwave and optical fields from system 1 to system 2, while too strong decay rates are not beneficial for generating entanglement. From Fig.3(b) and Fig.3(e), it can be seen that the degree of cascaded enhancement gradually decreases as the magnon decay rate increases, which is due to the greater impact on generating entanglement. 

\begin{figure}[htbp]
\centering\includegraphics[width=13cm]{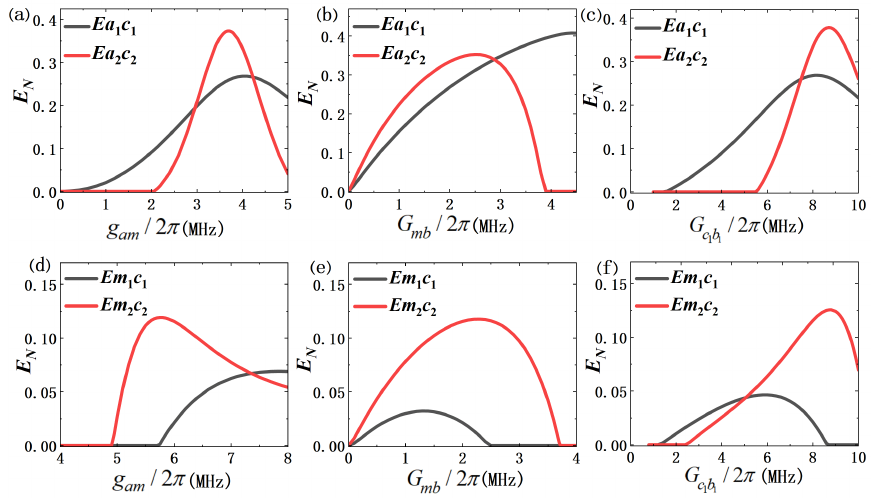}
\caption{The optomicrowave entanglement ($E_{a_1c_1}$ and $E_{a_2c_2}$) and the optomagnonic entanglement ($E_{m_1c_1}$ and $E_{m_2c_2}$) versus the microwave-magnon coupling strength $g_{am}$ (a,d), the effective magnonmechanical coupling strength $G_{mb}$ (b,e), and the effective optomechanical coupling strength $G_{c_1b_1}$ (c,f), respectively. All parameters being consistent with those in Fig.2, except the variables.}
\end{figure}
The variation of optomicrowave and optomagnonic entanglements as a function of the different coupling strengths for each system is illustrated in Fig.4. On the whole, cascaded enhancement of entanglement can be achieved within a certain range of coupling strengths. Both optomicrowave and optomagnonic entanglements exhibit a trend of "first increasing and then decreasing" with increasing coupling strengths. This behavior can be attributed to the fact that the increase in the coupling strength initially plays a positive role in the generation or transfer of entanglement, but it will become negative when the coupling strength increases to a certain extent. 
From the steady state solution of the optical mode, $G_{c_2b_2}=\frac{\sqrt{\eta_2}}{10}\frac{g_{c_2b_2}}{g_{c_1b_1}} G_{c_1b_1}$ can be obtained, which means that the value of $G_{c_2b_2}$ is constrained by the ratio of the single-photon optomechanical coupling strength in system 2 to that in system 1. Therefore, the variation of optomicrowave and optomagnonic entanglements in system 2 as a function of this ratio is investigated, as shown in Fig.5. It can be found from Fig.5 that the entanglement exhibits a trend of "first increasing and then decreasing" as $g_{c_{2}b_{2}}$ increases. This behavior can be understood as: with increasing the ratio $g_{c_2b_2}/g_{c_1b_1}$, the optomechanical coupling in system 2 strengthens, which initially enhances the entanglement transferred to $E_{a_2c_2}$ and $E_{m_2c_2}$; however, when $g_{c_{2}b_{2}}$ reaches a certain value, the phonon mode primarily interacts with the optical mode, which weakens the interaction with the magnon mode and results in a decrease in the generated entanglement. In addition, the maximum entanglement occurs near $g_{c_{2}b_{2}}/g_{c_{1}b_{1}}=10$. For convenience, the ratio is set to $g_{c_2b_2}/g_{c_1b_1}=10$ in our scheme, to get a simplified relationship: $G_{c_2b_2}=\sqrt{\eta_2} G_{c_1b_1}$.
\begin{figure}[htbp]
\centering\includegraphics[width=10cm]{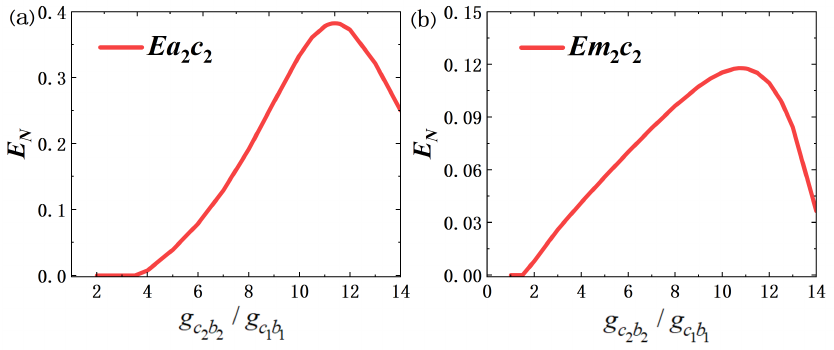}
\caption{The optomicrowave entanglement $E_{a_2c_2}$ (a) and the optomagnonic entanglement $E_{m_2c_2}$ (b) versus the ratio of the single-photon optomechanical coupling strengths $g_{c_2b_2}/g_{c_1b_1}$, respectively. Except for $G_{c_2b_2}$, the remaining parameters are consistent with those in Fig.2.}
\end{figure}

\begin{figure}[htbp]
\centering\includegraphics[width=13cm]{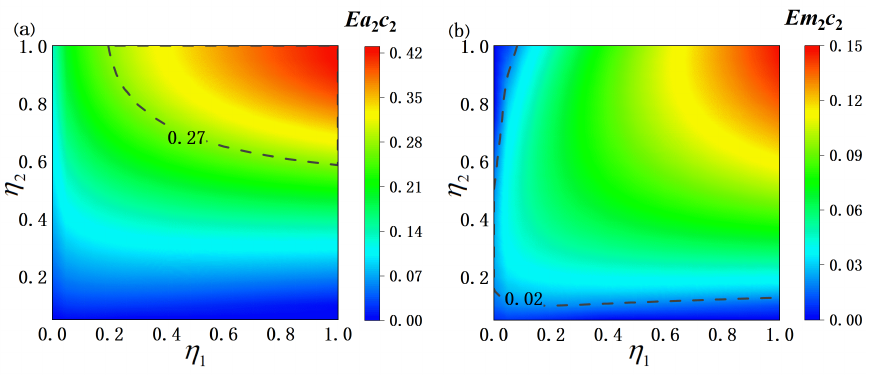}
\caption{The optomicrowave entanglement $E_{a_2c_2}$ (a) and the optomagnonic entanglement $E_{m_2c_2}$ (b) versus the transmission efficiency of microwave mode $\eta_1$ and the transmission efficiency of optical mode $\eta_2$. All parameters consistent with those in Fig.2, except variables.}
\end{figure}
The variation of optomicrowave and optomagnonic entanglements in system 2 as a function of microwave transmission efficiency $\eta_1$ and optical transmission efficiency $\eta_2$ is shown in Fig.6. The dashed lines (contour lines) in Fig.6(a) and Fig.6(b) show the results of the optomicrowave and the optomagnonic entanglements in system 1, respectively. It is obvious that the entanglements in system 2 increase with increasing transmission efficiency. Moreover, the enhancement of the optomicrowave entanglement in system 2 requires a higher transmission efficiency, while the enhancement of the optomagnonic entanglement has few requirements for transmission efficiency. This suggests that a higher transmission efficiency can strengthen the cascaded enhancement effect. Based on current experimental levels (the microwave transmission efficiency reaches $\eta=0.75$ \cite{170501,041052}), we choose $\eta_1=\eta_2=0.75$.

\begin{figure}[htbp]
\centering\includegraphics[width=13cm]{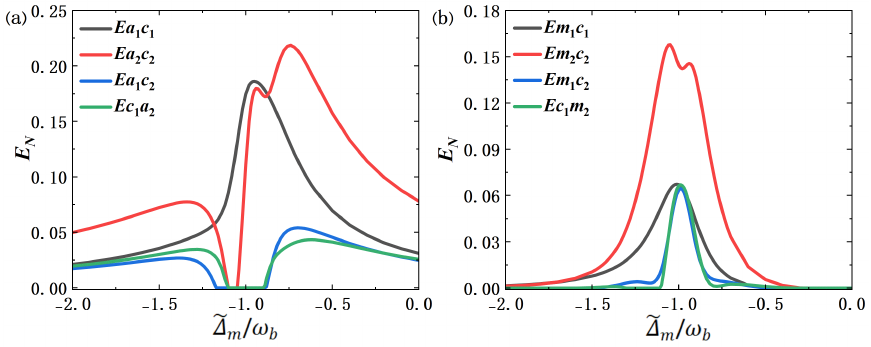}
\caption{The quadripartite optomicrowave entanglement (a) and the quadripartite optomagnonic entanglement (b) versus the effective detuning of magnon mode. (a) $\kappa_m=2\pi\times1 MHz$, $\kappa_c=2\pi\times3 MHz$, $G_{c_1b_1}=2\pi\times7 MHz$, $\Delta_a=-0.9\omega_b$; (b) $\kappa_c=\kappa_m=\kappa_a=2\pi\times1 MHz$. Other parameters are consistent with those in Fig.2.}
\end{figure}
If the two systems are considered as a whole, it is also possible to generate quadripartite optomicrowave and quadripartite optomagnonic entanglements, which are of particular interest. The variation of these two types of quadripartite entanglements as a function of effective detuning $\tilde{\Delta}_{m}$ is shown in Fig.7. It is found from Fig.7(a) that the quadripartite optomicrowave entanglement emerges in the region deviate from $\widetilde{\Delta}_m=-\omega_b$ and the optimal quadripartite optomicrowave entanglement appears near $\widetilde{\Delta}_m=-0.6\omega_b$. As shown in Fig.7(b), quadripartite optomagnonic entanglement is observed near $\widetilde{\Delta}_m=-\omega_b$ and the optimal quadripartite optomagnonic entanglement occurs at $\widetilde{\Delta}_m=-\omega_b$. Comparing these two figures, it can be concluded that quadripartite entanglement occurs mainly in the cascaded enhancement region of the entanglement. This indicates that the cascaded enhancement facilitates the generation of quadripartite entanglement. In the region where no entanglement enhancement occurs in Fig.7(a), the optomagnonic entanglement across system 1 and system 2 is zero, due to the weak correlation between the two systems. It is known that superconducting qubits and superconducting cavities running in the microwave frequency band are particularly suitable for quantum information processing tasks, whereas optical signals propagating through optical fibers are good at long-distance quantum information transmission. Therefore, the generation of multipartite optomicrowave entanglement is crucial for constructing hybrid quantum networks, enabling each subsystem to fully play its advantages. In addition, the preparation of multipartite optomagnonic entanglement provides a new thought for using light to control, design, detect, and transmit magnon's states.

\section{Conclusion}
We propose a scheme to enhance entanglement in optomicrowave and optomagnonic systems, as well as to prepare multipartite optomicrowave and optomagnonic entanglements, based on a cascaded optomagnomechanical system. In our analysis, we comprehensively examine the dependence of entanglement enhancement on various detunings, decay rates, and coupling strengths, providing the corresponding physical explanations and determing the optimal parameter ranges for entanglement enhancement, along with the basis of the chosen parameters. We also provide the basis of the chosen parameters. Furthermore, we also study the quadripartite optomicrowave and quadripartite optomagnonic entanglements for the entire cascaded system, analyzing the relationship between the generation of multipartite entanglement and the cascaded enhancement of entanglement, and identifying the parameter conditions for optimal quadripartite entanglement. The enhancement of optomicrowave entanglement and the preparation of multipartite optomicrowave entanglement are crucial for constructing hybrid quantum networks, allowing each subsystem to make full use of its advantages. The enhancement of optomagnonic entanglement and the preparation of multipartite optomagnonic entanglement offer a new approach for controlling, designing, detecting, and transmitting magnon's states with light, providing greater flexibility in manipulating and utilizing the quantum properties of magnon.

\begin{backmatter}
\bmsection{Funding}
National Key Research and Development Program of China (2021YFA1402002); Natural Science Foundation of China (NSFC) (11874249, 11874248, 11974225).
\bmsection{Disclosures}
The authors declare no conflicts of interest.
\bmsection{Data availability} Data underlying the results presented in this paper are not publicly available at this time but may be obtained from the authors upon reasonable request.
\end{backmatter}

\bibliography{ref}

\begin{thebibliography}{10}
\newcommand{\enquote}[1]{``#1''}

\bibitem{419326112}
G.~Kurizki, P.~Bertet, Y.~Kubo, K.~Mølmer, D.~Petrosyan, P.~Rabl, and J.~Schmiedmayer, \enquote{Quantum technologies with hybrid systems,} {\protect\JournalTitle{Proceedings of the National Academy of Sciences}} \textbf{112}, 3866--3873 (2015).

\bibitem{Clerk2020HybridQS}
A.~A. Clerk, K.~W. Lehnert, K.~W. Lehnert, P.~Bertet, J.~R. Petta, Y.~Nakamura, and Y.~Nakamura, \enquote{Hybrid quantum systems with circuit quantum electrodynamics,} {\protect\JournalTitle{Nature Physics}} \textbf{16}, 257--267 (2020).

\bibitem{Pirandola2016PhysicsUT}
S.~Pirandola and S.~L. Braunstein, \enquote{Physics: Unite to build a quantum internet,} {\protect\JournalTitle{Nature}} \textbf{532}, 169--171 (2016).

\bibitem{062620}
A.~P. Singh, K.~Sugisaki, S.~Prasannaa, B.~K. Sahoo, B.~P. Das, and Y.~Nakamura, \enquote{Experimental computations of atomic properties on a superconducting quantum processor,} {\protect\JournalTitle{Phys. Rev. A}} \textbf{110}, 062620 (2024).

\bibitem{Ladd2010QuantumC}
T.~D. Ladd, F.~Jelezko, R.~Laflamme, Y.~Nakamura, C.~R. Monroe, and J.~L. O'Brien, \enquote{Quantum computers,} {\protect\JournalTitle{Nature}} \textbf{464}, 45--53 (2010).

\bibitem{Du23}
P.~Du, Y.~Wang, K.~Liu, R.~Yang, and J.~Zhang, \enquote{Generation of large-scale continuous-variable cluster states multiplexed both in time and frequency domains,} {\protect\JournalTitle{Opt. Express}} \textbf{31}, 7535--7544 (2023).

\bibitem{153}
I.~M. Georgescu, S.~Ashhab, and F.~Nori, \enquote{Quantum simulation,} {\protect\JournalTitle{Rev. Mod. Phys.}} \textbf{86}, 153--185 (2014).

\bibitem{Arute2019QuantumSU}
F.~Arute, K.~Arya, R.~Babbush, D.~Bacon, J.~C. Bardin, R.~Barends, R.~Biswas, S.~Boixo, F.~G. S.~L. Brand{\~a}o, D.~A. Buell, B.~Burkett, Y.~Chen, Z.~Chen, B.~Chiaro, R.~Collins, W.~Courtney, A.~Dunsworth, E.~Farhi, B.~Foxen, A.~G. Fowler, C.~Gidney, M.~Giustina, R.~Graff, K.~Guerin, S.~Habegger, M.~P. Harrigan, M.~J. Hartmann, A.~K. Ho, M.~Hoffmann, T.~Huang, T.~Humble, S.~V. Isakov, E.~Jeffrey, Z.~Jiang, D.~Kafri, K.~Kechedzhi, J.~Kelly, P.~V. Klimov, S.~Knysh, A.~N. Korotkov, F.~Kostritsa, D.~Landhuis, M.~Lindmark, E.~Lucero, D.~I. Lyakh, S.~Mandr{\`a}, J.~R. McClean, M.~J. McEwen, A.~Megrant, X.~Mi, K.~Michielsen, M.~Mohseni, J.~Mutus, O.~Naaman, M.~Neeley, C.~J. Neill, M.~Y. Niu, E.~P. Ostby, A.~Petukhov, J.~C. Platt, C.~Quintana, E.~G. Rieffel, P.~Roushan, N.~C. Rubin, D.~T. Sank, K.~J. Satzinger, V.~N. Smelyanskiy, K.~J. Sung, M.~D. Trevithick, A.~Vainsencher, B.~Villalonga, T.~White, Z.~Yao, P.~Yeh, A.~Zalcman, H.~Neven, and J.~M. Martinis, \enquote{Quantum supremacy using a programmable
  superconducting processor,} {\protect\JournalTitle{Nature}} \textbf{574}, 505 -- 510 (2019).

\bibitem{260502}
P.~Magnard, S.~Storz, P.~Kurpiers, J.~Sch\"ar, F.~Marxer, J.~L\"utolf, T.~Walter, J.-C. Besse, M.~Gabureac, K.~Reuer, A.~Akin, B.~Royer, A.~Blais, and A.~Wallraff, \enquote{Microwave quantum link between superconducting circuits housed in spatially separated cryogenic systems,} {\protect\JournalTitle{Phys. Rev. Lett.}} \textbf{125}, 260502 (2020).

\bibitem{Xavier2019QuantumIP}
G.~B. Xavier and G.~Lima, \enquote{Quantum information processing with space-division multiplexing optical fibres,} {\protect\JournalTitle{Communications Physics}} \textbf{3}, 1--11 (2019).

\bibitem{Fan2022MicrowaveOpticsEV}
Z.~Fan, L.~Qiu, S.~Groblacher, and J.~Li, \enquote{Microwave‐optics entanglement via cavity optomagnomechanics,} {\protect\JournalTitle{Laser \& Photonics Reviews}} \textbf{17} (2022).

\bibitem{Luo23}
Y.-X. Luo, L.-J. Cong, Z.-G. Zheng, H.-Y. Liu, Y.~Ming, and R.-C. Yang, \enquote{Entanglement enhancement and epr steering based on a pt-symmetric-like cavity-opto-magnomechanical hybrid system,} {\protect\JournalTitle{Opt. Express}} \textbf{31}, 34764--34778 (2023).

\bibitem{Di23}
K.~Di, S.~Tan, L.~Wang, A.~Cheng, X.~Wang, Y.~Liu, and J.~Du, \enquote{High-efficiency entanglement of microwave fields in cavity opto-magnomechanical systems,} {\protect\JournalTitle{Opt. Express}} \textbf{31}, 29491--29503 (2023).

\bibitem{023501}
Z.-Y. Fan, H.~Qian, X.~Zuo, and J.~Li, \enquote{Entangling ferrimagnetic magnons with an atomic ensemble via optomagnomechanics,} {\protect\JournalTitle{Phys. Rev. A}} \textbf{108}, 023501 (2023).

\bibitem{Xin17}
J.~Xin, J.~Qi, and J.~Jing, \enquote{Enhancement of entanglement using cascaded four-wave mixing processes,} {\protect\JournalTitle{Opt. Lett.}} \textbf{42}, 366--369 (2017).

\bibitem{040305}
Z.~Yan, X.~Jia, X.~Su, Z.~Duan, C.~Xie, and K.~Peng, \enquote{Cascaded entanglement enhancement,} {\protect\JournalTitle{Phys. Rev. A}} \textbf{85}, 040305 (2012).

\bibitem{031201}
X.~Zuo, Z.-Y. Fan, H.~Qian, M.-S. Ding, H.~Tan, H.~Xiong, and J.~Li, \enquote{Cavity magnomechanics: from classical to quantum,} {\protect\JournalTitle{New J. Phys.}} \textbf{26}, 031201 (2024).

\bibitem{0370}
B.~{Zare Rameshti}, S.~{Viola Kusminskiy}, J.~A. Haigh, K.~Usami, D.~Lachance-Quirion, Y.~Nakamura, C.-M. Hu, H.~X. Tang, G.~E. Bauer, and Y.~M. Blanter, \enquote{Cavity magnonics,} {\protect\JournalTitle{Physics Reports}} \textbf{979}, 1--61 (2022). Cavity Magnonics.

\bibitem{070101}
D.~Lachance-Quirion, Y.~Tabuchi, A.~Gloppe, K.~Usami, and Y.~Nakamura, \enquote{Hybrid quantum systems based on magnonics,} {\protect\JournalTitle{Appl. Phys. Express}} \textbf{12}, 070101 (2019).

\bibitem{030405}
D.~Vitali, S.~Gigan, A.~Ferreira, H.~R. B\"ohm, P.~Tombesi, A.~Guerreiro, V.~Vedral, A.~Zeilinger, and M.~Aspelmeyer, \enquote{Optomechanical entanglement between a movable mirror and a cavity field,} {\protect\JournalTitle{Phys. Rev. Lett.}} \textbf{98}, 030405 (2007).

\bibitem{30005}
X.~Liu, R.~Yang, J.~Zhang, and T.~Zhang, \enquote{Generation of multipartite entangled states based on a double-longitudinal-mode cavity optomechanical system,} {\protect\JournalTitle{Opt. Express}} \textbf{31}, 30005--30019 (2023).

\bibitem{29581}
T.~Wang, L.~Wang, Y.~Liu, C.-H. Bai, D.-Y. Wang, H.~Wang, and S.~Zhang, \enquote{Temperature-resistant generation of robust entanglement with blue-detuning driving and mechanical gain.} {\protect\JournalTitle{Opt. Express}} \textbf{27 21}, 29581--29593 (2019).

\bibitem{203601}
J.~Li, S.-Y. Zhu, and G.~S. Agarwal, \enquote{Magnon-photon-phonon entanglement in cavity magnomechanics,} {\protect\JournalTitle{Phys. Rev. Lett.}} \textbf{121}, 203601 (2018).

\bibitem{013192}
H.~Tan and J.~Li, \enquote{Einstein-podolsky-rosen entanglement and asymmetric steering between distant macroscopic mechanical and magnonic systems,} {\protect\JournalTitle{Phys. Rev. Res.}} \textbf{3}, 013192 (2021).

\bibitem{032314}
G.~Vidal and R.~F. Werner, \enquote{Computable measure of entanglement,} {\protect\JournalTitle{Phys. Rev. A}} \textbf{65}, 032314 (2002).

\bibitem{090503}
M.~B. Plenio, \enquote{Logarithmic negativity: A full entanglement monotone that is not convex,} {\protect\JournalTitle{Phys. Rev. Lett.}} \textbf{95}, 090503 (2005).

\bibitem{170501}
N.~Roch, M.~E. Schwartz, F.~Motzoi, C.~Macklin, R.~Vijay, A.~W. Eddins, A.~N. Korotkov, K.~B. Whaley, M.~Sarovar, and I.~Siddiqi, \enquote{Observation of measurement-induced entanglement and quantum trajectories of remote superconducting qubits,} {\protect\JournalTitle{Phys. Rev. Lett.}} \textbf{112}, 170501 (2014).

\bibitem{041052}
A.~Chantasri, M.~E. Kimchi-Schwartz, N.~Roch, I.~Siddiqi, and A.~N. Jordan, \enquote{Quantum trajectories and their statistics for remotely entangled quantum bits,} {\protect\JournalTitle{Phys. Rev. X}} \textbf{6}, 041052 (2016).

\end{thebibliography}

\end{document}